\batchmode
\makeatletter
\def\input@path{{"/home/jacob/Documents/Work/My Papers/2026-The Historical Debate over the Ontological Wave Function/"}}
\makeatother
\documentclass[11pt,twoside,english]{article}
\usepackage[T1]{fontenc}
\usepackage[latin9]{inputenc}
\usepackage{color}
\definecolor{lyxboxbgcolor}{rgb}{0.980469, 0.941406, 0.902344}
\usepackage{babel}
\usepackage{amsmath}
\usepackage{amssymb}
\usepackage{geometry}
\usepackage{setspace}
\usepackage[pdftex,pdfusetitle,
 bookmarks=true,bookmarksnumbered=true,bookmarksopen=false,
 breaklinks=false,pdfborder={0 0 0},pdfborderstyle={},backref=false,colorlinks=false]
 {hyperref}

\makeatletter
\let\originalleft\left
\let\originalright\right
\renewcommand{\left}{\mathopen{}\mathclose\bgroup\originalleft}
\renewcommand{\right}{\aftergroup\egroup\originalright}

\def\smalloverbrace#1{\mathop{\vbox{\m@th\ialign{##\crcr%
      \noalign{\kern3\p@}%
      \tiny\downbracefill\crcr\noalign{\kern3\p@\nointerlineskip}%
      $\hfil\displaystyle{#1}\hfil$\crcr}}}\limits}

\def\smallunderbrace#1{\mathop{\vtop{\m@th\ialign{##\crcr
   $\hfil\displaystyle{#1}\hfil$\crcr
   \noalign{\kern3\p@\nointerlineskip}%
   \tiny\upbracefill\crcr\noalign{\kern3\p@}}}}\limits}

\DeclareMathAlphabet{\mymathbb}{U}{bbold}{m}{n}

\makeatother

\begin{document}
\title{Historical Debates over the Physical Reality of the Wave Function}
\author{Jacob A. Barandes\thanks{Departments of Philosophy and Physics, Harvard University, Cambridge, MA 02138; jacob\_barandes@harvard.edu; ORCID: 0000-0002-3740-4418}
\thanks{This manuscript is the accepted version, as published in \emph{The European Physical Journal H} (DOI:10.1140/epjh/s13129-026-00129-x).}}
\date{}

\maketitle

\begin{abstract}
This paper provides a detailed and unified historical account of
early debates over wave-function realism, the modern term for the
view that the wave function of quantum theory is physically real.
As this paper will show, the idea of physical waves associated with
particles had its roots in work by Einstein and de Broglie, who both
originally thought of these waves as propagating in three-dimensional
physical space. De Broglie quickly turned this wave\textendash particle
duality into an early pilot-wave theory, on which a particle's associated
``phase wave'' piloted or guided the particle along its trajectory.
Schrödinger built on de Broglie's phase-wave hypothesis to provide
a comprehensive account of the nascent quantum theory. However, Schrödinger's
new ``undulatory mechanics'' came at the cost of replacing de Broglie's
phase waves propagating in physical space with a wave function propagating
in a system's abstract configuration space. The present work will
argue that this move from three-dimensional physical space to a many-dimensional
configuration space was a decisive reason why the founders of quantum
theory uniformly abandoned the physical reality of the wave function.
This paper will further clarify that de Broglie introduced two distinct
pilot-wave theories, and will then argue that it was Bohm's rediscovery
of the second of these two pilot-wave theories over two decades later,
as well as Bohm's vociferous defense of wave-function realism, that
were responsible for resurrecting the idea of an ontological wave
function. Finally, this paper will describe how the influence of
Schrödinger and Bohm ended up playing a central role in Everett's
development of the ``many worlds'' interpretation.
\end{abstract}

\begin{center}
\global\long\def\quote#1{``#1"}%
\global\long\def\apostrophe{\textrm{'}}%
\global\long\def\slot{\phantom{x}}%
\global\long\def\eval#1{\left.#1\right\vert }%
\global\long\def\keyeq#1{\boxed{#1}}%
\global\long\def\importanteq#1{\boxed{\boxed{#1}}}%
\global\long\def\given{\vert}%
\global\long\def\mapping#1#2#3{#1:#2\to#3}%
\global\long\def\composition{\circ}%
\global\long\def\set#1{\left\{  #1\right\}  }%
\global\long\def\setindexed#1#2{\left\{  #1\right\}  _{#2}}%

\global\long\def\setbuild#1#2{\left\{  \left.\!#1\,\right|\,#2\right\}  }%
\global\long\def\complem{\mathrm{c}}%

\global\long\def\union{\cup}%
\global\long\def\intersection{\cap}%
\global\long\def\cartesianprod{\times}%
\global\long\def\disjointunion{\sqcup}%

\global\long\def\isomorphic{\cong}%

\global\long\def\setsize#1{\left|#1\right|}%
\global\long\def\defeq{\equiv}%
\global\long\def\conj{\ast}%
\global\long\def\overconj#1{\overline{#1}}%
\global\long\def\re{\mathrm{Re\,}}%
\global\long\def\im{\mathrm{Im\,}}%

\global\long\def\transp{\mathrm{T}}%
\global\long\def\tr{\mathrm{tr}}%
\global\long\def\adj{\dagger}%
\global\long\def\diag#1{\mathrm{diag}\left(#1\right)}%
\global\long\def\dotprod{\cdot}%
\global\long\def\crossprod{\times}%
\global\long\def\Probability#1{\mathrm{Prob}\left(#1\right)}%
\global\long\def\Amplitude#1{\mathrm{Amp}\left(#1\right)}%
\global\long\def\cov{\mathrm{cov}}%
\global\long\def\corr{\mathrm{corr}}%

\global\long\def\absval#1{\left\vert #1\right\vert }%
\global\long\def\expectval#1{\left\langle #1\right\rangle }%
\global\long\def\op#1{\hat{#1}}%

\global\long\def\bra#1{\left\langle #1\right|}%
\global\long\def\ket#1{\left|#1\right\rangle }%
\global\long\def\braket#1#2{\left\langle \left.\!#1\right|#2\right\rangle }%

\global\long\def\parens#1{(#1)}%
\global\long\def\bigparens#1{\big(#1\big)}%
\global\long\def\Bigparens#1{\Big(#1\Big)}%
\global\long\def\biggparens#1{\bigg(#1\bigg)}%
\global\long\def\Biggparens#1{\Bigg(#1\Bigg)}%
\global\long\def\bracks#1{[#1]}%
\global\long\def\bigbracks#1{\big[#1\big]}%
\global\long\def\Bigbracks#1{\Big[#1\Big]}%
\global\long\def\biggbracks#1{\bigg[#1\bigg]}%
\global\long\def\Biggbracks#1{\Bigg[#1\Bigg]}%
\global\long\def\curlies#1{\{#1\}}%
\global\long\def\bigcurlies#1{\big\{#1\big\}}%
\global\long\def\Bigcurlies#1{\Big\{#1\Big\}}%
\global\long\def\biggcurlies#1{\bigg\{#1\bigg\}}%
\global\long\def\Biggcurlies#1{\Bigg\{#1\Bigg\}}%
\global\long\def\verts#1{\vert#1\vert}%
\global\long\def\bigverts#1{\big\vert#1\big\vert}%
\global\long\def\Bigverts#1{\Big\vert#1\Big\vert}%
\global\long\def\biggverts#1{\bigg\vert#1\bigg\vert}%
\global\long\def\Biggverts#1{\Bigg\vert#1\Bigg\vert}%
\global\long\def\Verts#1{\Vert#1\Vert}%
\global\long\def\bigVerts#1{\big\Vert#1\big\Vert}%
\global\long\def\BigVerts#1{\Big\Vert#1\Big\Vert}%
\global\long\def\biggVerts#1{\bigg\Vert#1\bigg\Vert}%
\global\long\def\BiggVerts#1{\Bigg\Vert#1\Bigg\Vert}%
\global\long\def\ket#1{\vert#1\rangle}%
\global\long\def\bigket#1{\big\vert#1\big\rangle}%
\global\long\def\Bigket#1{\Big\vert#1\Big\rangle}%
\global\long\def\biggket#1{\bigg\vert#1\bigg\rangle}%
\global\long\def\Biggket#1{\Bigg\vert#1\Bigg\rangle}%
\global\long\def\bra#1{\langle#1\vert}%
\global\long\def\bigbra#1{\big\langle#1\big\vert}%
\global\long\def\Bigbra#1{\Big\langle#1\Big\vert}%
\global\long\def\biggbra#1{\bigg\langle#1\bigg\vert}%
\global\long\def\Biggbra#1{\Bigg\langle#1\Bigg\vert}%
\global\long\def\braket#1#2{\langle#1\vert#2\rangle}%
\global\long\def\bigbraket#1#2{\big\langle#1\big\vert#2\big\rangle}%
\global\long\def\Bigbraket#1#2{\Big\langle#1\Big\vert#2\Big\rangle}%
\global\long\def\biggbraket#1#2{\bigg\langle#1\bigg\vert#2\bigg\rangle}%
\global\long\def\Biggbraket#1#2{\Bigg\langle#1\Bigg\vert#2\Bigg\rangle}%
\global\long\def\angs#1{\langle#1\rangle}%
\global\long\def\bigangs#1{\big\langle#1\big\rangle}%
\global\long\def\Bigangs#1{\Big\langle#1\Big\rangle}%
\global\long\def\biggangs#1{\bigg\langle#1\bigg\rangle}%
\global\long\def\Biggangs#1{\Bigg\langle#1\Bigg\rangle}%

\global\long\def\vec#1{\mathbf{#1}}%
\global\long\def\vecgreek#1{\boldsymbol{#1}}%
\global\long\def\idmatrix{\mymathbb{1}}%
\global\long\def\projector{P}%
\global\long\def\permutationmatrix{\Sigma}%
\global\long\def\densitymatrix{\rho}%
\global\long\def\krausmatrix{K}%
\global\long\def\stochasticmatrix{\Gamma}%
\global\long\def\lindbladmatrix{L}%
\global\long\def\dynop{\Theta}%
\global\long\def\timeevop{U}%
\global\long\def\hadamardprod{\odot}%
\global\long\def\tensorprod{\otimes}%

\global\long\def\inprod#1#2{\left\langle #1,#2\right\rangle }%
\global\long\def\normket#1{\left\Vert #1\right\Vert }%
\global\long\def\hilbspace{\mathcal{H}}%
\global\long\def\samplespace{\Omega}%
\global\long\def\configspace{\mathcal{C}}%
\global\long\def\phasespace{\mathcal{P}}%
\global\long\def\spectrum{\sigma}%
\global\long\def\restrict#1#2{\left.#1\right\vert _{#2}}%
\global\long\def\from{\leftarrow}%
\global\long\def\statemap{\omega}%
\global\long\def\degangle#1{#1^{\circ}}%
\global\long\def\trivialvector{\tilde{v}}%
\global\long\def\eqsbrace#1{\left.#1\qquad\right\}  }%
\global\long\def\operator#1{\operatorname{#1}}%
\par\end{center}

\section{Introduction\label{sec:Introduction}}

In many pedagogical and formal treatments of quantum theory today,
the wave function plays a central, starring role. One is then confronted
with an obvious question: is the wave function physically real? That
is, is the wave function ontological?

Before proceeding, it will be important to disambiguate the term \textquoteleft wave
function.\textquoteright{} On the Dirac\textendash von Neumann formulation
of quantum theory (Dirac 1930, von Neumann 1932)\nocite{Dirac:1930pofm,vonNeumann:1932mgdq},
there is an important distinction to be made between, on the one hand,
an abstract quantum state and, on the other hand, a configuration-space
wave function. In the research literature, the term \textquoteleft wave
function,\textquoteright{} without qualification, is sometimes used
to refer to the former, and sometimes to the latter.

On the Dirac\textendash von Neumann axioms, an abstract quantum state
is represented by a vector $\ket{\Psi}$ in (or, more generally, an
operator $\rho$ on) a Hilbert space. By contrast, a configuration-space
wave function $\Psi\left(q\right)$ is defined as the set of complex-valued
components of a state vector $\ket{\Psi}$ with respect to a specific
orthonormal basis for the Hilbert space, where the members $\ket q$
of that orthonormal basis are indexed or labeled by a suitable set
of configurations $q$ making up a configuration space:\footnote{Technically speaking, if the configuration space is continuous, and
one is using a so-called \textquoteleft regular\textquoteright{} representation
of the appropriate Weyl algebra, then the basis vectors $\ket q$
will not be normalizable and will not actually belong to the Hilbert
space, but will instead be defined in terms of a suitable rigged Hilbert
space. These technicalities will not be important for the purposes
of this paper.} 
\begin{equation}
\ket{\Psi}=\int dq\,\Psi\left(q\right)\ket q,\qquad\Psi\left(q\right)=\braket q{\Psi}.\label{eq:DefConfigSpaceWaveFuncFromStateVec}
\end{equation}
The present work will be focused primarily on configuration-space
wave functions $\Psi\left(q\right)$.

The view that configuration-space wave functions are physically real
or ontological\textemdash or at least directly represent something
physically real or ontological\textemdash has come to be known as
\textquoteleft wave-function realism.\textquoteright{} In recent decades,
wave-function realism has generated a tremendous amount of scholarship
and controversy (Albert 1996; Albert, Loewer 1996; Lewis 2004; Ney,
Albert 2013; Myrvold 2015; Chen 2019; Wallace 2020; Ney 2021; Ney
2023)\nocite{Albert:1996eqm,AlbertLoewer:1996tossc,Lewis:2004lics,NeyAlbert:2013twfeotmoqm,Myrvold:2015wiaw,Chen:2019ratwf,Wallace:2020awr,Ney:2021twitwfamfqp,Ney:2023tafwfr}.

The purpose of the present work is not to wade into these contemporary
debates, but to trace their origins to discussions between the founders
of quantum theory. In particular, this paper will argue for the following
key points:
\begin{itemize}
\item The original wave\textendash particle duality introduced by Albert
Einstein in 1905 and further developed by Louis de Broglie from 1923
to 1926 involved waves propagating in three-dimensional physical space. 
\item On de Broglie's ``phase wave'' theory, the waves that accompanied
material particles also determined the motion of those particles according
to an early form of a \textquoteleft guiding equation.\textquoteright{}
Even then, de Broglie questioned the physical reality of his phase
waves because, on his theory, they propagated with superluminal phase
velocity.
\item Erwin Schrödinger, in his foundational papers on ``undulatory''\footnote{``Undulatory'' comes from the Latin root \emph{unda}, which means
``wave,'' and is also the source of the French word ``onde'' for
``wave.''} or ``wave mechanics'' in 1926 and 1927, moved de Broglie's waves
from three-dimensional physical space to the much more abstract, many-dimensional
configuration space of the system in question. Schrödinger immediately
tried to grapple with the physical reality of his newly introduced
wave function, to the point of introducing a proto-Everettian (``many
worlds'') interpretation. Schrödinger quickly (albeit temporarily)
settled on the nuanced view that his wave function was an ontological
object whose observable effect was to determine distributions of electric
charge in three-dimensional physical space.
\item In many letters to colleagues in 1926 and 1927, including in his famous
``God does not play dice'' letter to Max Born in 1926, Einstein
repeatedly and vocally expressed his opposition to the physicality
of waves in a many-dimensional configuration space. 
\item Born's 1926 statistical interpretation of the wave function complicated
discussions over the physicality of the configuration-space wave function.
On this statistical interpretation, the modulus-square of the wave
function gave the probability density with which a measurement of
the system's configuration would yield a specific result.
\item Starting with a presentation to the fifth Solvay Conference in 1927,
and building on the work of Schrödinger and Born, de Broglie developed
two distinct pilot-wave theories on which quantum-mechanical waves
piloted or guided particles along their trajectories in three-dimensional
physical space.
\item In a 1927 paper, de Broglie gave a detailed presentation of the first
of these two pilot-wave theories, which eventually became known as
his ``double-solution theory.'' This pilot-wave theory was based
on two conceptually distinct kinds of waves that both satisfied the
same wave equation. The first kind of wave conjecturally featured
solitonic singularities or local maxima that de Broglie identified
with material particles. The second kind of wave realized Born's statistical
interpretation. De Broglie found the mathematics of this double-solution
theory to be quite complicated, and he eventually set it aside.
\item In a lecture in 1928, Schrödinger publicly disavowed his earlier
view that his wave function in a configuration space should be understood
to be physically real.
\item De Broglie presented a second, conceptually distinct pilot-wave theory
in a 1930 book. This second pilot-wave theory involved only one kind
of wave\textemdash Schrödinger's wave function. On this theory, the
wave function piloted or guided material particles along their trajectories
according to a precise and general guiding equation, where de Broglie
treated those material particles as a separate form of ontology rather
than as solitonic singularities derived from the wave function itself.
In the later chapters of that book, de Broglie attempted to make sense
of this pilot-wave theory in the multi-particle case, in which the
configuration space was many-dimensional. De Broglie decided that
a wave function propagating in such a many-dimensional space could
only be treated as abstract and fictitious, and not as ontological.
\item As a result, after completing his 1930 book, de Broglie abandoned
both of his pilot-wave theories for several decades, only returning
to his double-solution theory after David Bohm contacted him in 1951
to share his own pilot-wave theory with de Broglie.
\item A denial of the physical reality of Schrödinger's wave function was
one of the few things that the major figures in the development of
quantum theory\textemdash Einstein, de Broglie, Schrödinger, Born,
Heisenberg, Bohr, Pauli, and Dirac\textemdash all agreed on.
\item Bohm introduced the first comprehensive treatment of decoherence in
the later chapters of his 1951 textbook on quantum theory, a contribution
to the progress of physics that has not been sufficiently recognized.\footnote{Most references credit Dieter Zeh and Wojciech Zurek for the discovery
of decoherence, including the Stanford Encyclopedia of Philosophy's
entry ``The Role of Decoherence in Quantum Mechanics'' (Bacciagaluppi
2025)\nocite{Bacciagaluppi:2025trodiqm}. Indeed, the front matter
in a 2025 book by Zurek himself says ``Wojciech Hubert Zurek is credited
with developing the theory of decoherence'' (Zurek 2025)\nocite{Zurek:2025daqdfqftcr}.
Until February 2024, the Wikipedia entry ``Quantum decoherence''
did not mention Bohm's pivotal contributions (Wikipedia 2026)\nocite{Wikipediacontributors:2025qd}.
A marvelous exception is a 2004 paper by Brown and Wallace (Brown,
Wallace 2004, Footnote 9)\nocite{BrownWallace:2004smpdbble}.}
\item Bohm then developed what he initially thought was an original pilot-wave
theory from 1951 to 1952, eventually publishing a pair of foundational
papers laying out the theory in 1952. The first of these two papers
focused primarily on the axiomatic construction of Bohm's pilot-wave
theory, and the second paper used decoherence to explain how his pilot-wave
theory provided a means of resolving the measurement problem.
\item Bohm's pilot-wave theory was identical to de Broglie's second pilot-wave
theory, as Bohm acknowledged explicitly in correspondence with Wolfgang
Pauli in 1951. 
\item One of Bohm's major contributions was using decoherence to solve old
problems with de Broglie's version of the theory.
\item Another of Bohm's major contributions was to breathe new life into
what is now called \textquoteleft wave-function realism\textquoteright{}
by arguing that the pilot wave should be regarded as physically real
or ontological, despite the fact that it propagated in a many-dimensional
configuration space. Bohm mentioned this claim in only a limited way
in his two 1952 papers introducing his pilot-wave theory, but made
the case far more extensively and robustly in written correspondence
with Pauli in 1951.
\item Bohm's defense of wave-function realism inspired others to take wave
functions, and even more abstract quantum states, to be physically
real entities. In particular, Hugh Everett was clearly aware of Bohm's
pivotal development of the theory of decoherence, and wrote in detail
about Bohm's work in his unpublished, extended 1956 dissertation.
Going farther than Bohm, Everett took the quantum state of the universe,
treated now as an element of a Hilbert space, to be the sole ontology
of nature. Everett was also greatly influenced by Schrödinger's writings
in the early 1950s, in which Schrödinger attempted to resurrect his
realism about the wave function and cast doubt on the need for probability
axioms.
\end{itemize}

\section{The History of Waves in Quantum Theory\label{sec:The-History-of-Waves-in-Quantum-Theory}}

\subsection{De Broglie's Phase Waves\label{subsec:De-Broglie's-Phase-Waves}}

In the autumn of 1923, Louis de Broglie wrote a paper titled ``\emph{Ondes et Quanta}''
(``Waves and Quanta,'' de Broglie 1923a)\nocite{DeBroglie:1923oeq}
that was inspired both by Albert Einstein's 1905 introduction of quanta
of light (Einstein 1905)\nocite{Einstein:1905uedeuvdlbhg} and also
by an apparent conflict between special relativity and quantum theory. 

On the one hand, relative to a fixed observer in some inertial reference
frame, a microscopic object moving at a speed $v$, and thus traveling
at a fraction $\beta=v/c$ of the speed of light $c$, should exhibit
time dilation. As a consequence, any internal periodic process inside
the object with proper frequency $\nu_{0}$ should appear to decrease
to a \emph{lower} frequency 
\begin{equation}
\nu_{1}=\nu_{0}\sqrt{1-\beta^{2}}<\nu_{0}.\label{eq:deBroglieLowerFrequency}
\end{equation}

On the other hand, the proper mass\textendash energy $E_{0}=m_{0}c^{2}$
of the microscopic object should be related to some sort of proper
frequency $\nu_{0}$ according to Planck's quantum formula $E_{0}=h\nu_{0}$,
where $h$ is Planck's constant. If the object were in motion at a
fraction $\beta=v/c$ of the speed of light, then the mass\textendash energy
$E_{0}$ should Lorentz-transform to the \emph{larger} value $m_{0}c^{2}/\sqrt{1-\beta^{2}}>m_{0}c^{2}$,
and thus the proper frequency $\nu_{0}$ should Lorentz-transform
to the correspondingly \emph{higher} frequency 
\begin{equation}
\nu=\nu_{0}/\sqrt{1-\beta^{2}}>\nu_{0}.\label{eq:deBroglieHigherFrequency}
\end{equation}

To de Broglie, this disparity between the lower value \eqref{eq:deBroglieLowerFrequency}
of the internal periodic frequency $\nu_{1}$ and the higher value
\eqref{eq:deBroglieHigherFrequency} of the Planck frequency $\nu$
suggested that the Planck frequency did not refer to a periodic process
internal to the microscopic object. De Broglie posited instead that
the Planck frequency referred to some kind of wave associated with
the object, propagating along with the object in three-dimensional
physical space. If this wave shared the same proper frequency $\nu_{0}$
as some internal periodic process, then because of the discrepancy
$\nu>\nu_{1}$, these two frequencies could only \emph{remain} synchronized
or in harmony for nonzero speeds $\beta\ne0$ if the wave had a faster
phase velocity than the physical velocity of the microscopic object.
Intuitively speaking, the wave's higher phase velocity could then
\textquoteleft spread out\textquoteright{} the peaks and valleys of
the wave to keep them in harmony with the peaks and valleys of the
object's internal periodic process.

A short calculation showed that in order to maintain phase harmony
(``\emph{théorème de l\textquoteright harmonie des phases}'' or
``\emph{l\textquoteright accord des phases}''), the wave would
need a phase velocity of $c/\beta=c^{2}/v>c$, meaning that the wave
needed to have a phase velocity faster than light. De Broglie concluded
that this ``phase wave'' (``\emph{onde de phase}'') could not
transport energy, and was therefore a ``fictitious wave'' (``\emph{onde fictive}''): 
\begin{quotation}
Suppose now that at the time $t=0$, the moving object coincides in
space with a wave of frequency $\nu$ defined above and propagating
in the same direction as it with the speed $c/\beta$. This wave,
with a speed greater than $c$, cannot correspond to transport of
energy; we will consider it only as a fictitious wave associated with
the motion of the object. {[}De Broglie 1923a, in literal English
translation from the original French by the author of the present
work{]}\nocite{DeBroglie:1923oeq}
\end{quotation}
Nonetheless, de Broglie used these propagating phase waves to try
to account for various quantization formulas that were known from
early work on quantum theory.

In a follow-up paper in the same year, titled ``\emph{Quanta de Lumière, Diffraction et Interfèrences}''
(``Quanta of Light, Diffraction and Interference,'' de Broglie 1923b)\nocite{DeBroglie:1923qdldei},
de Broglie also argued that on this phase-wave hypothesis, each particle
of matter moved at a velocity equal to the \emph{group} velocity of
its associated wave packet, and along a trajectory that matched the
geometric ray of that associated wave packet. Indeed, in that second
paper, de Broglie had already begun to use the language of ``guiding''
(``\emph{guidant}'') for his phase waves: 
\begin{quotation}
We therefore conceive of the phase wave as guiding the displacements
of energy, and this is what can permit the synthesis of undulations
and quanta. {[}De Broglie 1923b, in literal English translation from
the original French by the author of the present work{]}\nocite{DeBroglie:1923qdldei}
\end{quotation}
Both in that paper, and in a subsequent paper written in English also
in the same year, de Broglie referred to his phase waves as ``non-material''
(``\emph{non matèrielle}''), again due to their superluminal propagation
speeds (de Broglie 1923b, 1923c)\nocite{DeBroglie:1923qdldei,DeBroglie:1923waq}.

De Broglie expanded on these ideas in his 1924 doctoral dissertation,
titled ``\emph{Recherches sur la théorie des Quanta}'' (``Research
on the Quantum Theory,'' de Broglie 1924)\nocite{deBroglie:1924rsltdq},
for which he would go on to win the 1929 Nobel Prize in Physics. In
Chapter 2, Section V of de Broglie's dissertation, he presented an
early version of a \textquoteleft guiding equation\textquoteright{}
in the form 
\begin{equation}
O_{i}=\frac{1}{h}J_{i},\label{eq:EarlydeBrogleGuidingEqFromThesis}
\end{equation}
 where, following de Broglie's notation, $i$ is a Lorentz index. 

On the right-hand side of \eqref{eq:EarlydeBrogleGuidingEqFromThesis},
$J_{i}$ was the object's ``world vector'' (``\emph{vecteur d\textquoteright Univers}''),
\begin{equation}
J_{i}=m_{0}cu_{i}+e\varphi_{i}.\label{eq:deBroglieWorldVector}
\end{equation}
 This formula involved the object's proper or rest mass $m_{0}$,
the speed of light $c$, and the object's dimensionless four-velocity
$u_{i}=dx_{i}/ds$, where $ds=\sqrt{c^{2}{dt}^{2}-{dx}^{2}-{dy}^{2}-{dz}^{2}}$
denoted infinitesimal increments along the object's worldline. This
formula for $J_{i}$ also included the object's electric charge $e$
and the electromagnetic gauge potentials $\varphi_{i}$ (not to be
confused with the notation $\varphi$ for phase functions to be introduced
shortly). 

On the left-hand side of \eqref{eq:EarlydeBrogleGuidingEqFromThesis},
$O_{i}$ was the phase wave's four-dimensional ``wave world vector''
(``\emph{le vecteur Onde d\textquoteright Univers}''), given in
terms of the phase function $\varphi\left(x,y,z,t\right)$ of the
phase wave according to the differential relationship 
\begin{equation}
d\varphi=2\pi\sum_{i}O_{i}dx^{i}.\label{eq:deBroglieWaveVectorAsSum}
\end{equation}
 On comparing this equation with the multivariable chain rule, which
takes the form 
\begin{equation}
d\varphi=\sum_{i}\frac{\partial\varphi}{\partial x^{i}}dx^{i},\label{eq:deBroglieWaveVectorChainRule}
\end{equation}
 one reads off the four individual equations 
\begin{equation}
\frac{\partial\varphi}{\partial x^{i}}=2\pi O_{i},\label{eq:deBroglieWaveVectorPartialDerivatives}
\end{equation}
 which are equivalent to 
\begin{equation}
O_{i}=\frac{1}{2\pi}\frac{\partial\varphi}{\partial x^{i}}.\label{eq:deBroglieWaveFectorAsPartials}
\end{equation}
Identifying the partial derivatives on the right-hand side as the
components of the gradient $\nabla\varphi$ of the phase function
$\varphi\left(x,y,z,t\right)$, one then immediately sees that the
spatial components $\vec O=\left(O_{x},O_{y},O_{z}\right)$ of $O_{i}$
are given by the gradient of $\varphi$, up to a reciprocal factor
of $2\pi$: 
\begin{equation}
\vec O=\frac{1}{2\pi}\nabla\varphi.\label{eq:deBroglieWaveVectorAsGradient}
\end{equation}
 The guiding equation \eqref{eq:EarlydeBrogleGuidingEqFromThesis}
then has spatial parts given by 
\begin{equation}
\frac{1}{2\pi}\nabla\varphi=\frac{1}{h}\vec J,\label{eq:EarlydeBroglieGuidingEquationIntermed}
\end{equation}
 so that 
\begin{equation}
\vec J=\left(\frac{h}{2\pi}\right)\nabla\varphi.\label{eq:EarlydeBroglieGuidingEquationRearranged}
\end{equation}

In the non-relativistic limit $v\ll c$, and in the absence of electromagnetic
fields, one has $c\vec u\approx\vec v$, where $\vec v$ is the particle's
ordinary velocity, and $\varphi_{i}=0$. The spatial parts of the
world vector \eqref{eq:deBroglieWorldVector} then approximately become
\[
\vec J\approx m_{0}\vec v,
\]
 and thus the spatial parts of the guiding equation \eqref{eq:EarlydeBrogleGuidingEqFromThesis}
reduce to 
\begin{equation}
\vec v=\frac{1}{m_{0}}\left(\frac{h}{2\pi}\right)\nabla\varphi.\label{eq:DeBroglieNonRelativisticGuidingEq}
\end{equation}
 Here the quantity appearing in parentheses, $h/2\pi$, would now
be called the reduced Planck constant $\hbar$ (\textquoteleft h-bar\textquoteright ),
a notation introduced by Paul Dirac a few years later (Dirac 1928)\nocite{Dirac:1928zqde}.

The equation \eqref{eq:DeBroglieNonRelativisticGuidingEq} will turn
out to be closely related to the guiding equation of a pilot-wave
theory. However, it is worth keeping in mind that when de Broglie
wrote his dissertation, his phase-wave theory only applied in the
geometrical-optics limit, meaning for very short wavelengths, and
therefore under circumstances in which he could directly relate Maupertuis's
principle for the paths of material particles to Fermat's principle
for the paths of light rays. At that point in his work, de Broglie
had not yet clearly specified a detailed physical relationship between
particles of matter and his phase waves more generally.

One saw the first steps toward describing such a physical relationship
in some of de Broglie's writings starting in 1926. For example, in
a 1926 paper titled ``\emph{Les Principes de la Nouvelle Mécanique Ondulatoire}''
(``The Principles of the New Wave Mechanics,'' de Broglie 1926)\nocite{DeBroglie:1926lpdlnmo},
in Section 10 (``\emph{Le point matériel}''), de Broglie wrote
that in the geometrical-optics limit, a particle of matter could be
identified with ``a very pronounced maximum'' (``\emph{un maximum très pronuncé}'')
in its associated phase wave. He then added: 
\begin{quotation}
The rays of the associated wave are the possible trajectories of the
moving body. But in a \emph{given} motion, one of the rays is of particular
physical importance, namely, the ray which is actually described by
the moving body. The associated wave, in so far as we have imagined
it up to the present, does not enable us to understand what distinguishes
this ray from the others. In order to get some light on this point
it is necessary, as I pointed out in my thesis, to consider not only
a monochromatic associated wave, but a group of associated waves with
frequencies very close to each other. {[}De Broglie 1926, translated
from the French to English by W.~M. Deans, emphasis in the original{]}\nocite{DeBroglie:1926lpdlnmo}
\end{quotation}
 Later in that 1926 paper, de Broglie briefly returned to this wave-derived
conception of particles of matter, referring to each now as a ``point
singularity'' (``\emph{singularité ponctuelle}'') in its associated
phase wave, but then also casting doubt on the idea outside the geometrical-optics
limit: 
\begin{quotation}
If in this case there still exists in every group of waves a \emph{point}-singularity
worthy of the name of material particle, we can easily conceive that
each of the groups of waves is propagated according to a law like
(59), where the coefficients depend on the position at that time of
the singularities in the other groups; but if, in this kind of motion,
there are no longer any well-defined material particles, the question
becomes obscure. {[}Ibid., emphasis in the original{]}
\end{quotation}

\subsection{Schrödinger's Wave Function\label{subsec:Schr=0000F6dinger's-Wave-Function}}

In early 1926, motivated in part by de Broglie's phase-wave hypothesis,
and also by classical Hamilton\textendash Jacobi theory, Erwin Schrödinger
introduced a new kind of wave $\psi\left(q,t\right)$. This was his
famous ``wave function'' (``\emph{Wellenfunktion}''), where $q$
labels an abstract point in the system's classical configuration space
and where $t$, as usual, is the time. 

Schrödinger laid out the foundations of his new theory of ``undulatory''
or ``wave mechanics'' in a series of four German-language papers,
all sharing the title ``\emph{Quantisierung als Eigenwertproblem}''
(``Quantization as an Eigenvalue Problem,'' Schrödinger 1926a\textendash d)\nocite{Schrodinger:1926qae1,Schrodinger:1926qae2,Schrodinger:1926qae3,Schrodinger:1926qae4}.
In the first of these four \emph{Eigenwertproblem} papers, Schrödinger
began with the restricted Hamilton\textendash Jacobi equation for
a system with classical Hamiltonian $H$ and total energy $E$: 
\begin{equation}
H\left(q,\frac{\partial S}{\partial q}\right)=E.\label{eq:SchrodingerHamiltonJacobiRestricted}
\end{equation}
 Then Schrödinger introduced $\psi$ according to 
\begin{equation}
S=K\lg\psi,\label{eq:SchrodingerFirstDefinitionWaveFunction}
\end{equation}
 where $\lg$ was his notation for the natural logarithm and where
$K$ was a constant with units of action. Schrödinger would eventually
identify $K$ as 
\begin{equation}
K=\frac{h}{2\pi},\label{eq:SchrodingerKConst}
\end{equation}
 which, again, as in \eqref{eq:DeBroglieNonRelativisticGuidingEq},
would now be called the reduced Planck constant $\hbar$.

According to Schrödinger's new undulatory theory, the wave function
evolved with time according to what is now known as the configuration-space
Schrödinger equation, which Schrödinger originally wrote down as his
eq. ($4^{\prime\prime}$) in the fourth \emph{Eigenwertproblem} paper
(Schrödinger 1926d)\nocite{Schrodinger:1926qae4}:\footnote{See also eq. (32) of Schrödinger (1926e)\nocite{Schrodinger:1926autotmoaam}.}
\begin{equation}
\Delta\psi-\frac{8\pi^{2}}{h^{2}}V\psi\mp\frac{4\pi i}{h}\frac{\partial\psi}{\partial t}=0.\label{eq:SchrodingerOriginalEq}
\end{equation}
 In this equation, $\Delta$ was a second-order differential operator
acting on the system's configuration space that included the inertial
masses of the various particles, and $V$ was an arbitrary potential
function on the configuration space. The ambiguous sign in the third
term reflected the freedom to choose either $\psi$ or its complex
conjugate $\bar{\psi}$ as the function satisfying the equation. In
terms of the reduced Planck constant $\hbar=h/2\pi$, and after some
algebraic rearrangement, this equation takes the more familiar form
\begin{equation}
\mp i\hbar\frac{\partial\psi}{\partial t}=-\frac{\hbar^{2}}{2}\Delta\psi+V\psi,\label{eq:SchrodingerOriginalEqMoreFamiliarForm}
\end{equation}
 where the positive sign convention is generally used today. Schrödinger
would go on to share the 1933 Nobel Prize in Physics for this work,
alongside Dirac.

An immediate question was the physical status of the wave function.
Schrödinger wavered on this question. In the first \emph{Eigenwertproblem}
paper, Schrödinger wrote:\footnote{All English translations of Schrödinger's foundational papers on wave
mechanics here are due to J.~F. Shearer and W.~M. Deans (1928)\nocite{Schrodinger:1928cpowm}.} 
\begin{quotation}
It is, of course, strongly suggested that we should try to connect
the function $\psi$ with some \emph{vibration process} in the atom,
which would more nearly approach reality than the electronic orbits,
the real existence of which is being very much questioned today. {[}Schrödinger
1926a, translated from the German to English by Shearer and Deans,
emphasis in the original{]}\nocite{Schrodinger:1926qae1}
\end{quotation}
In the second \emph{Eigenwertproblem} paper, Schrödinger referred
explicitly to $\psi$ ``as a physical quantity'' (``\emph{physikalische Größe}''),
and added that, as a consequence of its physicality, ``the function
$\psi$ must be single-valued, finite, and continuous throughout configuration
space'' (Schrödinger 1926b)\nocite{Schrodinger:1926qae2}.

As was transparently evident from the Schrödinger equation \eqref{eq:SchrodingerOriginalEqMoreFamiliarForm},
Schrödinger's wave function was complex-valued, rather than real-valued,
as Schrödinger complained in a letter to Hendrik Lorentz, dated June
6, 1926: 
\begin{quotation}
What is unpleasant here, and indeed directly to be objected to, is
the use of complex numbers. $\psi$ is surely fundamentally a real
function {[}...{]}.'' {[}Schrödinger 1963, translated from the German
to English by M.~J. Klein{]}\nocite{Schrodinger:1963lthal6j1}
\end{quotation}
In the fourth \emph{Eigenwertproblem} paper, Schrödinger wrote:
\begin{quotation}
Meantime, there is no doubt a certain crudeness in the use of a \emph{complex}
wave function. If it were unavoidable \emph{in principle}, and not
merely a facilitation of the calculation, this would mean that there
are in principle \emph{two} wave functions, which must be used \emph{together}
in order to obtain information on the state of the system. This somewhat
unacceptable inference admits, I believe, of the very much more congenial
interpretation that the state of the system is given by a real function
and its time-derivative. Our inability to give more accurate information
about this is intimately connected with the fact that, in the pair
of equations ($4^{\prime\prime}$), we have before us only the \emph{substitute}\textemdash extraordinarily
convenient for the calculation, to be sure\textemdash for a real wave
equation of probably the fourth order, which, however, I have not
succeeded in forming for the non-conservative case. {[}Schrödinger
1926d, translated from the German to English, emphasis in the original{]}\nocite{Schrodinger:1926qae4}
\end{quotation}

Schrödinger was also troubled by the fact that his wave function $\psi\left(q,t\right)$
was generically defined not in three-dimensional physical space, but
in the abstract, many-dimensional configuration space, or ``$q$-space,''
of the system in question. Hendrik Lorentz raised this concern in
a letter that he sent to Schrödinger, dated May 27, 1926:
\begin{quotation}
If I had to choose now between your wave mechanics and the matrix
mechanics {[}of Born, Heisenberg, and Jordan{]}, I would give the
preference to the former, because of its greater intuitive clarity,
so long as one only has to deal with the three coordinates $x,y,z$.
If, however, there are more degrees of freedom, then I cannot interpret
the waves and vibrations physically, and I must therefore decide in
favor of matrix mechanics. But your way of thinking has the advantage
for this case too that it brings us closer to the real solution of
the equations; the eigenvalue problem is the same in principle for
a higher dimensional $q$-space as it is for three dimensional space.
{[}Schrödinger 1963, translated from the German to English by M.~J.
Klein{]}\nocite{Schrodinger:1963lthal6j1}
\end{quotation}
Schrödinger himself thought extensively about this issue as well.
In a separate paper published in 1926, titled ``\emph{Über das Verhältnis der Heisenberg\textendash Born\textendash Jordanschen
Quantenmechanik zu der meinen}'' (``On the Relationship Between Heisenberg\textendash Born\textendash Jordan
Quantum Mechanics and My Own,'' Schrödinger 1926f)\nocite{Schrodinger:1926udvdhbjqzdm},
Schrödinger wrote: 
\begin{quotation}
The difficulty in regard to the problem of \emph{several} electrons,
which mainly lies in the fact that $\psi$ is a function in \emph{configuration
space}, not in real space, must be mentioned. {[}Schrödinger 1926f,
translated from the German to English, emphasis in the original{]}\nocite{Schrodinger:1926udvdhbjqzdm}
\end{quotation}
One saw further concern from Schrödinger in a paper that he published
in English at the end of 1926, titled ``An Undulatory Theory of the
Mechanics of Atoms and Molecules'': 
\begin{quotation}
This {[}the quantity $\psi\bar{\psi}${]} of course, like $\psi$
itself, is in the general case a function of the generalized coordinates
$q_{1},\dots,q_{N}$ and the time,\textemdash not a function of ordinary
space and time as in ordinary wave-problems. This raises some difficulty
in attaching a physical meaning to the wave-function. {[}Schrödinger
1926e{]}\nocite{Schrodinger:1926autotmoaam}
\end{quotation}
In his June 6, 1926 reply to the May 27 letter from Lorentz mentioned
above, Schrödinger expressed a more elaborate interpretation of his
configuration-space wave function: 
\begin{quotation}
You mention the difficulty of projecting the waves in $q$-space,
when there are more than three coordinates, into ordinary three dimensional
space and of interpreting them physically there. I have been very
sensitive to this difficulty for a long time but believe that I have
now overcome it. I believe (and I have worked it out at the end of
the third article) that the physical meaning belongs not to the quantity
itself but rather to a \emph{quadratic} function of it. \emph{There}
I chose the real part of $\psi\bar{\psi}$, where $\psi$ is taken
to be complex in the obvious way (for criticism, see below) and the
bar denotes the complex conjugate. \emph{Now} I want to choose more
simply $\psi\bar{\psi}$, that is, the square of the absolute value
of the quantity $\psi$. If we now have to deal with $N$ particles,
then $\psi\bar{\psi}$ (just as $\psi$ itself) is a function of $3N$
variables or, as I want to say, of $N$ three dimensional spaces,
$R_{1},R_{2},\dots,R_{N}$. Now first let $R_{1}$ be identified with
the real space and integrate $\psi\bar{\psi}$ over $R_{2},\dots,R_{N}$;
second, identify $R_{2}$ with the real space and integrate over $R_{1},R_{3},\dots,R_{N}$;
and so on. The $N$ individual results are to be added after they
have been multiplied by certain constants which characterize the particles
(their charges, according to the former theory). I consider the result
to be the electric charge density in real space. In this manner one
obtains for an atom with many electrons exactly what Born\textendash Heisenberg\textendash Jordan
designate as the transition probability, with the new and plausible
meaning ``component of the electric moment'' (strictly speaking
\emph{that} partial moment which oscillates with the \emph{emission}
frequency in question). {[}Schrödinger 1963, translated from the German
to English by M.~J. Klein, emphasis in the original{]}\nocite{Schrodinger:1963lthal6j1}
\end{quotation}
One finds the clearest and most extensive statement of Schrödinger's
early views on his wave function in the fourth \emph{Eigenwertproblem}
paper from 1926. This statement is likewise worth quoting in full: 
\begin{quotation}
This rule is now equivalent to the following conception, which allows
the true meaning of $\psi$ to stand out more clearly. $\psi\bar{\psi}$
is a kind of \emph{weight-function} in the system's configuration
space. The \emph{wave-mechanical} configuration of the system is a
\emph{superposition} of many, strictly speaking of \emph{all}, point-mechanical
configurations kinematically possible. Thus, each point-mechanical
configuration contributes to the true wave-mechanical configuration
with a certain \emph{weight}, which is given precisely by $\psi\bar{\psi}$.
If we like paradoxes, we may say that the system exists, as it were,
simultaneously in all the positions kinematically imaginable, but
not ``equally strongly'' in all.\footnote{Note the proto-Everettian ``many worlds'' interpretation here. The
present work will return to the connection between Schrödinger and
Everett in Section~\ref{sec:Conclusion}.} In macroscopic motions, the weight-function is practically concentrated
in a small region of positions, which are practically indistinguishable.
The centre of gravity of this region in configuration space travels
over distances which are macroscopically perceptible. In problems
of microscopic motions, we are in any case interested \emph{also},
and in certain cases even \emph{mainly}, in the varying \emph{distribution}
over the region.

This new interpretation may shock us at first glance, since we have
often previously spoken in such an intuitive concrete way of the ``$\psi$-vibrations''
as though of something quite real. But there is something tangibly
real behind the present conception also, namely, the very real electrodynamically
effective fluctuations of the electric space-density. The $\psi$-function
is to do no more and no less than permit of the totality of these
fluctuations being mastered and surveyed mathematically by a single
partial differential equation. We have repeatedly called attention
to the fact that the $\psi$-function itself cannot and may not be
interpreted directly in terms of three-dimensional space\textemdash however
much the one-electron problem tends to mislead us on this point\textemdash because
it is in general a function in configuration space, not real space.
{[}Schrödinger 1926d, translated from the German to English, emphasis
in the original{]}\nocite{Schrodinger:1926qae4}
\end{quotation}

Schrödinger's configuration-space wave function did not sit well with
Albert Einstein. In a series of letters to Hendrik Lorentz, Paul Ehrenfest,
and Arnold Sommerfeld through 1926 and 1927 (Howard 1990)\nocite{Howard:1990nskwnsdotpoe1eewatqmocs},
Einstein expressed his dismay and his disbelief repeatedly:
\begin{itemize}
\item ``Schrödinger's conception of the quantum rules makes a great impression
on me; it seems to me to be a bit of reality, however unclear the
sense of waves in $n$-dimensional $q$-space remains.'' {[}In a
letter dated May 1, 1926, to Lorentz{]}
\item ``Schrödinger's works are wonderful\textemdash but even so one nevertheless
hardly comes closer to a real understanding. The field in a many-dimensional
coordinate space does not smell like something real.'' {[}In a letter
dated June 18, 1926, to Ehrenfest{]}
\item ``The method of Schrödinger seems indeed more correctly conceived
than that of Heisenberg, and yet it is hard to place a function in
coordinate space and view it as an equivalent for a motion. But if
one could succeed in doing something similar in four-dimensional space,
then it would be more satisfying.'' {[}In a letter dated June 22,
1926, to Lorentz again{]}
\item ``Of the new attempts to obtain a deeper formulation of the quantum
laws, that by Schrödinger pleases me most. If only the undulatory
fields introduced there could be transplanted from the $n$-dimensional
coordinate space to the 3 or 4 dimensional! The Heisenberg\textendash Dirac
theories compel my admiration, but to me they don't smell like reality.''
{[}In a letter dated August 21, 1926, to Sommerfeld{]}
\item ``Admiringly\textemdash mistrustfully I stand opposed to quantum
mechanics. I do not understand Dirac at all in points of detail (Compton
effect) . . . . Schrödinger is, in the beginning, very captivating.
But the waves in $n$-dimensional coordinate space are indigestible,
as well as the absence of any understanding of the frequency of the
emitted light.'' {[}In a letter dated August 28, 1926, to Ehrenfest
again{]}
\item ``The quantum theory has been completely Schrödingerized and has
much practical success from that. But this can nevertheless not be
the description of a real process. It is a mystery.'' {[}In a letter
dated February 16, 1927, to Lorentz again{]}
\end{itemize}
Of all these letters, the most famous was surely Einstein's ``God
does not play dice'' letter to Max Born, dated December 4, 1926:\footnote{In the original German: 
\begin{quotation}
Die Quantenmechanik ist sehr achtung-gebietend. Aber eine innere Stimme
sagt mir, daß das doch nicht der wahre Jakob ist. Die Theorie liefert
viel, aber dem Geheimnis des Alten bringt sie uns kaum näher. Jedenfalls
bin ich überzeugt, daß \emph{der} nicht würfelt. Wellen im $3n$-dimensionalen
Raum, deren Geschwindigkeit durch potentielle Energie (z. B. Gummibänder)
reguliert wird...
\end{quotation}
}
\begin{quotation}
Quantum mechanics is very awe-inspiring. But an inner voice tells
me that this is not the real Jacob after all.\footnote{\emph{Der Wahre Jacob} was the title of a German satirical magazine
in circulation at various points from 1879 to 1933. } The theory delivers a lot, but to the secrets of the Old One it brings
us hardly any closer. Anyway I am convinced that \emph{He} does not
roll dice. Waves in $3n$-dimensional space\footnote{In the canonical English translation by Irene Born (1971)\nocite{EinsteinBornBorn:1971tbelcbaeamahbf1t1wcbmb},
the $n$ is missing in ``3$n$-dimensional space'' (``\emph{$3n$-dimensionalen Raum}''),
obscuring Einstein's concerns about the physical reality of waves
in a many-dimensional configuration space.} whose speed through potential energy (e.g. rubber bands) is regulated...
{[}Translated from the German to English by the author of the present
work, emphasis in the original{]}
\end{quotation}

It is interesting to note that Einstein proposed a hidden-variables
theory of his own, in the spring of 1927, and therefore chronologically
after all the critical correspondence presented above. As recounted
by Belousek (1996)\nocite{Belousek:1996e1uhtibcas}, Einstein read
a draft of his paper at a meeting of the Prussian Academy of Sciences
in Berlin on May 5, 1927. His paper was titled ``Does Schrödinger\textquoteright s
Wave Mechanics Determine the Motion of a System Completely or Only
in the Sense of Statistics?'' (``\emph{Bestimmt Schrödingers Wellenmechanik die Bewegung eines Systems vollständig
oder nur im Sinne der Statistik}?'') Einstein's proposed theory was noncommittal about the physical
meaning of the wave function, using it merely as a means of assigning
definite motions to particles. As Belousek noted, however, Einstein
identified a problem with his theory in how it handled composite systems
consisting of independent subsystems. Einstein ultimately withdrew
his paper just weeks later, before it appeared in print. Einstein
was not forthcoming about his reasons for withdrawing his paper, but
Belousek, in particular, has speculated that in addition to the problem
with composite systems, Einstein may also have been motivated by his
concerns over the physicality of wave functions in many-dimensional
configuration spaces. As Del Santo (2019)\nocite{Santo:2019sfrnfdhmoeab}
has emphasized, there is strong historical evidence that these sorts
of issues surrounding realism were far more pressing to Einstein than
questions about determinism.

De Broglie was not any more amenable to the idea of ontological wave
functions propagating in configuration spaces than Einstein was. In
the 1926 paper discussed earlier, ``\emph{Les Principes de la Nouvelle Mécanique Ondulatoire},''
de Broglie explicitly mentioned Schrödinger's recent work: 
\begin{quotation}
But what happens if the dynamical phenomenon runs its course in a
domain of the order of a wave-length, as in the case of the atom?
Schrödinger thinks that then we can no longer speak of a material
particle describing a trajectory. {[}...{]} This is a difficult problem,
which we shall meet with in various forms farther on. {[}De Broglie
1926, translated from the French to English by W.~M. Deans{]}\nocite{DeBroglie:1926lpdlnmo}
\end{quotation}
 De Broglie then described Schrödinger's view, on which each ensemble
of $N$ particles should be described by a wave function propagating
in the system's $3N$-dimensional configuration space, rather than
in three-dimensional physical space: 
\begin{quotation}
Schrödinger has leanings toward this latter opinion, and, in order
to establish the dynamics of systems, he generalizes my idea of phase-waves:
with the motion of the system of $N$ moving bodies he associates
a wave in the space of $3N$ dimensions imagined by the classical
theories in order to represent the motion of the whole system by the
displacement of a single representative point. This wave associated
with the system of $N$ particles would thus be a function of $3N$
space-coordinates and of the time. {[}Ibid.{]}
\end{quotation}
 Continuing, de Broglie expressly rejected the ontological reality
of Schrödinger's wave function: 
\begin{quotation}
Up to the present I have been unable to accept this point of view;
for me the associated waves have physical reality and must be expressed
as functions of the three space-co-ordinates and of the time. I cannot
dwell further on this difficult question; it is obvious, from the
above, that the wave dynamics of systems does not seem to be solidly
established as yet. {[}Ibid.{]} 
\end{quotation}

At the end of a series of lectures on wave mechanics that Schrödinger
delivered in English at the Royal Institution in London in March 1928,
he recapitulated his 1926 interpretation of the wave function in Section
15 (``The physical meaning of the generalized $\psi$-function''): 
\begin{quotation}
Perhaps the latter conclusions are obscured by the fact that we have
hitherto avoided putting forward any definite assumption as to the
physical interpretation of the function $\psi\left(q_{1},q_{2},\dots,q_{n},t\right)$
relating to a system whose configuration in terms of ordinary mechanics
is described by the generalized co-ordinates $q_{1},q_{2},\dots,q_{n}$.
This interpretation \emph{is} a very delicate question. As an obvious
generalization of the procedure of spreading out the electronic charge
according to a relative density function $\psi\bar{\psi}$ (which
furnished satisfactory results in the one-electron problem; see section
5), the following view would present itself in the case of a general
mechanical system: the real natural system does not behave like the
picture which ordinary mechanics forms of it (e.g., a system of point-charges
in a definite configuration), but rather behaves like what would be
the result of spreading out the system, described by $q_{1},\dots,q_{n}$,
\emph{throughout its configuration-space} in accordance with a relative
density function $\psi\bar{\psi}$. This would mean that, if the ordinary
mechanical picture is to be made use of at all, the actual system
behaves like the ordinary mechanical picture, present in all its possible
configurations at the same time, though ``stronger'' in some of
them than in others.\footnote{Notice again the proto-Everettian ``many worlds'' interpretation
that Schrödinger presented here.} {[}Schrödinger 1928, emphasis in the original{]}\nocite{Schrodinger:1928flowmdatrilo571a1m1}
\end{quotation}
However, Schrödinger then immediately described having had a change
of heart: 
\begin{quotation}
I maintained this view for some time. The fact that it proves \emph{very}
useful can be seen from the one-electron problem (see section 5).
\emph{No} other interpretation of the $\psi$-function is capable
of making us \emph{understand} the large amount of information which
the constants $a_{kl}$ furnish about the intensity and polarization
of the radiation. Yet this way of putting the matter is surely not
quite satisfactory. For what does the expression ``to behave like''
mean in the preceding sentences? The ``behavior'' of the $\psi$-function,
i.e. its development in time, is governed by nothing like the laws
of classical mechanics; it is governed by the wave-equation.\textemdash{}
{[}Ibid., emphasis in the original{]}
\end{quotation}
Schrödinger then mentioned Max Born's statistical interpretation,
which Born had included in a paper that he had originally submitted
for publication on June 25, 1926 (Born 1926)\nocite{Born:1926zqds}:\footnote{As is well known, in the main text of the paper, Born erroneously
identified the statistical probability with the wave function itself.
In a footnote that he added during the correction process, he wrote
that, on further reflection, it was not the wave function itself,
but its (modulus) square, that should be interpreted as the relevant
probability. Born would go on to win the 1954 Nobel Prize in Physics
for this realization.} 
\begin{quotation}
An obvious \emph{statistical} interpretation of the $\psi$-function
has been put forward, viz. that it does not relate to a single system
at all but to an assemblage of systems, $\psi\bar{\psi}$ determining
the fraction of the systems which happen to be in a definite configuration.
This view is a little unsatisfactory, since it offers no explanation
whatever why the quantities $a_{kl}$ yield all the information which
they do yield. {[}Ibid., emphasis in the original{]}\footnote{In the following portion of that lecture, Schrödinger also foreshadowed
John Bell's 1990 article ``Against \textquoteleft Measurement\textquoteright{}
'' (Bell 1990)\nocite{Bell:1990am} with a criticism of the vague
language of measurement:
\begin{quotation}
In connexion with the statistical interpretation it has been said
that to any physical quantity which would have a definite physical
meaning and be in principle (\emph{principiell}) measurable according
to the classical picture of the atom, there belong definite proper
values (just as e.g. the proper values $E_{k}$, belong to the energy);
and it has been said that the result of measuring such a quantity
will always be one or the other of these proper values, but never
anything intermediate. It seems to me that this statement contains
a rather vague conception, namely that of \emph{measuring} a quantity
(e.g. energy or moment of momentum), which relates to the classical
picture of the atom, i.e. to an obviously wrong one. {[}Schrödinger
1928, emphasis in the original{]}\nocite{Schrodinger:1928flowmdatrilo571a1m1}
\end{quotation}
}
\end{quotation}

It is worth also briefly mentioning Paul Dirac's early views on the
wave function, about which he said very little. An exception was the
first chapter of the first edition of his 1930 textbook, \emph{Principles of Quantum Mechanics},
in which Dirac took an explicitly instrumentalist view that dismissed
regarding the wave function as ontological:\footnote{Although Dirac rewrote this part of his book in later editions, he
did not present a conceptually or philosophically different view in
those later editions.} 
\begin{quotation}
To obtain a consistent theory of light which shall include interference
and diffraction phenomena, we must consider the photons as being controlled
by waves, in some way which cannot be understood from the point of
view of ordinary mechanics. This intimate connexion between waves
and particles is of very great generality in the new quantum mechanics.
It occurs not only in the case of light. \emph{All} particles are
connected in this way with waves, which control them and give rise
to interference and diffraction phenomena under suitable conditions.
The influence of the waves on the motion of the particles is less
noticeable the more massive the particles and only in the case of
photons, the lightest of all particles, is it easily demonstrated.

The waves and particles should be regarded as two abstractions which
are useful for describing the same physical reality. One must not
picture this reality as containing both the waves and particles together
and try to construct a mechanism, acting according to classical laws,
which shall correctly describe their connexion and account for the
motion of the particles. Any such attempt would be quite opposed to
the principles by which modern physics advances. What quantum mechanics
does is to try to formulate the underlying laws in such a way that
one can determine from them without ambiguity what will happen under
any given experimental conditions. It would be useless and meaningless
to attempt to go more deeply into the relations between waves and
particles than is required for this purpose.

{[}...{]}

The reader may, perhaps, feel that we have not really solved the difficulty
of the conflict between the waves and the corpuscles, but have merely
talked about it in a certain way and, by using some of the concepts
of waves and some of corpuscles, have arrived at a formal account
of the phenomena, which does not really tell us anything that we did
not know before. The difficulty of the conflict between the waves
and corpuscles is, however, actually solved as soon as one can give
an unambiguous answer to any experimental question. \emph{The only
object of theoretical physics is to calculate results that can be
compared with experiment}, and it is quite unnecessary that any satisfying
description of the whole course of the phenomena should be given.
{[}Dirac 1930, emphasis in the original{]}\nocite{Dirac:1930pofm}
\end{quotation}

\subsection{De Broglie's Pilot-Wave Theory\label{subsec:De-Broglie's-Pilot-Wave-Theory}}

De Broglie sought a notion of particles that made sense even beyond
the geometrical-optics limit. In a 1927 paper, titled ``\emph{La Mécanique Ondulatoire et la Structure Atomique de la Matière et
du Rayonnement}'' (``Wave Mechanics and the Atomic Structure of Matter and Radiation,''
de Broglie 1927)\nocite{DeBroglie:1927lmoelsadlmedr}, de Broglie
initiated a different research direction based on what he called ``the
principle of the double solution'' (``\emph{principe de la double solution}'').
This new approach featured a \emph{pair} of waves, both propagating
in three-dimensional physical space, and that both satisfied the same
appropriately specified wave equation. One of these waves was intended
to have a singularity in it representing a particle of matter, and
the other was continuous and represented the flow of probability.
As a matter of history, it was this paper that first featured the
guiding equation in its more modern-looking form \eqref{eq:DeBroglieNonRelativisticGuidingEq},
as de Broglie's eq. ($26^{\prime}$), 
\begin{equation}
\overrightarrow{v_{M}}=\frac{1}{m_{0}}\overrightarrow{\mathrm{grad}\varphi_{1}},\label{eq:deBroglieGuidingEquationForDoubleSolution}
\end{equation}
 where $\overrightarrow{v_{M}}$ was the velocity of the matter particle,
$m_{0}$ was its inertial mass, and $\varphi_{1}$ was the phase function
of de Broglie's singularity-bearing wave. De Broglie would put his
double-solution theory aside for many years, before returning to it
in the 1950s, a story that he retold in the very first article published
in the newly founded journal \emph{Foundations of Physics} in 1970,
titled ``The Reinterpretation of Wave Mechanics'' (de Broglie 1970)\nocite{DeBroglie:1970trowm}.

In the meantime, de Broglie laid out a conceptually distinct pilot-wave
theory in his 1930 book \emph{Introduction à l'Étude de la Mécanique Ondulatoire}
(\emph{An Introduction to the Study of Wave Mechanics}, de Broglie
1930)\nocite{DeBroglie:1930ialedlmo}. De Broglie's 1930 book made
no mention of his double-solution theory, and only glancingly mentioned
his earlier supposition that particles of matter were singularities
in phase waves.

In the introduction to the book, de Broglie laid out his detailed
views on his understanding of wave\textendash particle duality. He
began by describing why he could not support the notion that particles
were merely wave packets:\footnote{All English translations of de Broglie's 1930 book here are due to
H.~T. Flint (1930)\nocite{DeBroglie:1930ialedlmo}.}
\begin{quotation}
But what does this duality of waves and particles mean? This is a
very difficult question and one which is still far from being very
clearly elucidated.

The simplest idea is that which Schrödinger put forward at the beginning
of his work, viz. that the particle or the electron is constituted
by a group of waves; it is a \textquotedblleft wave packet.\textquotedblright{}
We have seen that this can be maintained so long as mechanical phenomena
are considered which are in harmony with the old dynamics, that is
to say in the new language, phenomena in which the propagation of
the associated wave obeys the laws of geometrical optics. Unfortunately,
when we pass to the domain proper of the new theory it appears scarcely
possible to support this idea which is so attractive on account of
its simplicity. In an experiment like that of the diffraction of an
electron by a crystal the wave packet would be completely dispersed
and destroyed; as a result no particles would be found in the diffracted
bundles. In other words, if they were simple wave packets the particles
would have no stable existence. {[}De Broglie 1930, ``General Introduction''{]}\nocite{DeBroglie:1930ialedlmo}
\end{quotation}
De Broglie then clarified his earlier view that particles could be
regarded as singularities in his phase waves, before explaining why
he no longer found that view tenable, either, and would not mention
it again in the book: 
\begin{quotation}
If it appears impossible to maintain Schrödinger\textquoteright s
view in all its consequences, neither is it easy to develop another
opinion with which the author has for a long time associated himself
and according to which the particle is a singularity in a wave phenomenon.
In the special case of the uniform motion of a particle it is possible
to find a solution of the wave equation showing a moving singularity
and capable of representing the particle. But it is very difficult
to make the generalisation to the case of non-uniform motion, and
there are serious objections to this point of view; we shall not discuss
this difficulty any further in this volume. {[}Ibid.{]}
\end{quotation}
De Broglie then turned to his new pilot-wave theory, which he first
explicitly presented at the fifth Solvay conference in 1927, and which
featured a pilot wave that guided his particles (``\emph{On peut donc concevoir le corpuscule comme guidé par l'onde qui joue
le rôle d'onde-pilote}''): 
\begin{quotation}
The author has also made another suggestion which is published in
his report to the Fifth Solvay Congress. We have seen that we must
always associate a wave with a particle, and so the idea which is
in best agreement with the older views of physics is to consider the
wave as a reality and as occupying a certain region of space, while
the particle is regarded as a material point having a definite position
in the wave. This is the basis of the suggestion. Since it is necessary,
as we have said above, that the intensity of the wave should be proportional
at each point to the probability of the occurrence of the particle
at that point, we must attempt to connect the motion of the particle
with the propagation of the wave so that this relation is automatically
realised in every case. It is, in fact, actually possible to establish
a connection between the motion of the particle and the propagation
of the wave, so that if at an initial instant the intensity of the
wave measures the required probability the same is true at all later
instants. We may thus suppose that the particle is guided by the wave
which plays the part of a pilot-wave. This view permits of an interesting
visualization of the corpuscular motion in wave mechanics without
too wide a departure from classical ideas. {[}Ibid.{]}
\end{quotation}
However, de Broglie then previewed his concerns with his pilot-wave
idea: 
\begin{quotation}
Unfortunately, we encounter very serious objections to this view also,
and these will be pointed out in the course of the book. It is not
possible to regard the theory of the pilot-wave as satisfactory. Nevertheless,
since the equations on which this theory rests are sound, we may preserve
some of its consequences by giving to it a modified form in agreement
with ideas developed independently by Kennard. Instead of speaking
of the motion and of the trajectory of the particles, we speak of
the motion and of the trajectory of the \textquotedblleft elements
of probability\textquotedblright{} and in this way the difficulties
noted are avoided. {[}Ibid., with reference to a paper by Kennard
(1928)\nocite{Kennard:1928otqmoasop}{]}
\end{quotation}
De Broglie then pointed out that the \textquoteleft Copenhagen\textquoteright{}
interpretation associated with Heisenberg and Bohr rejected an ontological
view of the wave function: 
\begin{quotation}
Finally, there is a fourth point of view developed by Heisenberg and
Bohr which is most favoured at present. This point of view is a little
disconcerting at first sight, but yet it appears to contain a large
body of truth. According to this view, the wave does not represent
a physical phenomenon taking place in a region of space; rather it
is simply a symbolic representation of what we know about the particle.
{[}Ibid.{]}
\end{quotation}

Putting all these concerns aside, in Section IX.V (``\emph{La théorie de l'onde-pilote}''),
de Broglie posited that the wave function itself, obeying a wave equation
like the Schrödinger equation \eqref{eq:SchrodingerOriginalEq}, acted
as a pilot wave that determined the velocities of de Broglie's particles
according to a guiding equation that generalized \eqref{eq:DeBroglieNonRelativisticGuidingEq}
and \eqref{eq:deBroglieGuidingEquationForDoubleSolution}:
\begin{quotation}
If now we still wish to retain the classical conception of the particle
in the domain proper of the new mechanics, that is to say, outside
the approximation to geometrical optics, we naturally wish to maintain
the identity of the particle with one of the probability elements
and to represent the state of affairs by imagining on the one hand
the wave, and on the other the particle to be localised in space,
and we connect the motion of the particle with the propagation of
the wave by the relation:\footnote{Note that de Broglie chose the negative sign convention in his version
of the Schrödinger equation \eqref{eq:SchrodingerOriginalEq}, which
de Broglie wrote as $\Delta\Psi-\left(8\pi^{2}m/h^{2}\right)F\Psi=\left(4\pi im/h\right)\partial\Psi/\partial t$
in his eq. (IX.1). This sign convention led to an additional minus
sign in his guiding equation $\overrightarrow{v}=-\left(1/m\right)\mathrm{grad}\varphi$
in his eqs. (IX.5) and (IX.38).} 
\begin{equation}
\overrightarrow{v}=-\frac{1}{m}\mathrm{grad}\varphi\qquad\left[\textrm{de Broglie's eq. }\left(\textrm{IX.38}\right)\right],\label{eq:deBroglieGuidingEquation1930Book}
\end{equation}
where $\varphi$ is the phase of the wave defined by {[}de Broglie's{]}
(32). The velocity of the particle is thus determined at each instant
if we know the initial position, and thus its path also is determined.
Moreover, from the formulae of the first paragraph, if we know the
form of the associated $\psi$-wave and if we know that initially
the probability of occurrence of the particle at a point is equal
to the intensity of the wave at the point, it will be so automatically
at every succeeding instant; thus the principle of interference will
be satisfied. We may describe this theory as the pilot-wave theory,
because we imagine the wave as guiding the motion of the particle.
{[}De Broglie 1930, Chapter IX, Section 5{]}\nocite{DeBroglie:1930ialedlmo}
\end{quotation}

Despite occasional comments to the contrary in the research literature
(for example, Dewdney et al. 1992)\nocite{DewdneyHortonLamMalikSchmidt:1992wpdatioqm},
de Broglie generalized his pilot-wave theory to the multi-particle
case in his 1930 book, as laid out in his Chapters XIV (``Wave Mechanics
of Systems of Particles'') and XV (``The Interpretation of the Wave
Associated with the Motion of a System''). In particular, in addressing
the multi-particle case, de Broglie began referring explicitly to
the system's configuration space (``\emph{l'espace de configuration}''),
which Flint translated into English as ``generalised space'':
\begin{quotation}
Here, as in the case of the single particle, we might be tempted to
develop a pilot-wave theory, at the same time maintaining that the
particles of the system have a definite position in space and that,
consequently, the representative point has also a definite position
at each instant in the generalised {[}i.e., configuration{]} space.
We should then admit that the representative point associated with
a $\psi$-wave has the motion defined by (15) {[}de Broglie's guiding
equation{]}, coinciding always with one of the probability elements.
{[}De Broglie 1930, Chapter XV, Section 4{]}\nocite{DeBroglie:1930ialedlmo}
\end{quotation}

Even as he was introducing his pilot-wave theory, however, de Broglie
was already expressing doubts about the reality of the pilot wave
as propagating in a configuration space, in contrast to waves propagating
in ordinary space (``\emph{l'espace ordinaire}''): 
\begin{quotation}
Moreover, we have, in agreement with Schrödinger's work, considered
wave propagation in the fictitious generalised {[}i.e., configuration{]}
space. It has not yet been possible to connect this propagation of
a single wave in a fictitious space with that of one or several waves
in ordinary space.

This impossibility seems to strengthen the view that no physical reality
is to be attached to the associated wave, but that it is simply a
symbolic representation of probability. {[}De Broglie 1930, Chapter
XIV (``Wave Mechanics of Systems of Particles''), Section 4{]}\nocite{DeBroglie:1930ialedlmo}
\end{quotation}
De Broglie returned to these concerns about multi-particle systems
later in the book: 
\begin{quotation}
Here, as in the case of the single particle, we might be tempted to
develop a pilot-wave theory, at the same time maintaining that the
particles of the system have a definite position in space and that,
consequently, the representative point has also a definite position
at each instant in the generalised {[}configuration{]} space.  {[}...{]}
Unfortunately all the difficulties arise once more which we discovered
in the pilot-wave theory of a {[}single{]} particle, and it is even
more difficult to consider the theory as offering an actual physical
picture of the phenomena because of the abstract and fictitious character
of the propagation of a wave in generalised {[}configuration{]} space.
{[}De Broglie 1930, Chapter XV (``The Interpretation of the Wave
Associated with the Motion of a System''), Section 4{]}\nocite{DeBroglie:1930ialedlmo} 
\end{quotation}
De Broglie echoed these concerns many years later, in a letter to
Wolfgang Pauli dated February 10, 1952:
\begin{quotation}
I became persuaded, upon reflection, of the ineffectiveness of the
pilot-wave theory as I had presented it to the Solvay Conference because
it has the particle ``guided'' by a wave $\psi$ that is merely
the representation of a probability and whose fictitious character
is evident (propagation in configuration space, in particular). {[}Pauli
1996, parenthetical statement in the original, in literal English
translation from the original French by the author of the present
work{]}\nocite{Pauli:1996scwbehao}
\end{quotation}

To see how de Broglie's construction worked in more modern notation
for a system of $N$ particles with configuration $q$ consisting
of $n=3N$ degrees of freedom $q_{1},\dots,q_{n}$, it will be useful
to write the Schrödinger equation for the configuration-space wave
function $\Psi\left(q,t\right)$ in the more schematic form\footnote{Note again that de Broglie chose the negative sign convention in his
version of the Schrödinger equation \eqref{eq:SchrodingerOriginalEq},
leading to an additional minus sign in his own version of the guiding
equation in his 1930 book.} 
\begin{equation}
i\,\partial_{t}\Psi=-\Delta\Psi+V\Psi,\label{eq:SchematicSchrodingerEq}
\end{equation}
 where $\partial_{t}=\partial/\partial t$ is the partial derivative
with respect to the time $t$, and where factors of $\hbar$ have
been suppressed by a suitable choice of measurement units. Suppose
that the second-order differential operator $\Delta$ appearing in
the Schrödinger equation \eqref{eq:SchematicSchrodingerEq} is expressible
in the schematic form 
\begin{equation}
\Delta=\sum_{i,j}\frac{1}{2}\partial_{i}\left(\mu_{ij}\partial_{j}\right),\label{eq:SchematicSecondOrderDifferentialOp}
\end{equation}
 where $\mu_{ij}=\mu_{ji}$ is a symmetric array of real-valued functions
of the coordinates (assumed for simplicity to have unit determinant),
and where $\partial_{i}=\partial/\partial q_{i}$ and $\partial_{j}=\partial/\partial q_{j}$
are differential operators acting along the directions of the respective
coordinates $q_{i}$ and $q_{j}$. Writing the complex-valued wave
function in polar form in terms of a real-valued radial function $R\left(q,t\right)$
and a real-valued phase function $\theta\left(q,t\right)$, 
\begin{equation}
\Psi\left(q,t\right)=R\left(q,t\right)e^{i\theta\left(q,t\right)},\label{eq:SchematicWaveFunctionPolarDecomposition}
\end{equation}
 one can define a guiding equation to consist of the statement that
the velocities $\dot{Q}_{i}\left(t\right)$ of the particles are given
by the gradients $\partial_{j}\theta$ of the phase function, evaluated
at the system's actual instantaneous configuration $Q\left(t\right)=\left(Q_{1}\left(t\right),\dots,Q_{n}\left(t\right)\right)$
at the time $t$:\footnote{Compare this equation for $\dot{q}_{i}$ with de Broglie's eq. (XV.15),
which he wrote as $q^{\prime}_{i}=-\sum_{k}\mu^{ik}\partial\varphi/\partial q_{k}$
(de Broglie 1930)\nocite{DeBroglie:1930ialedlmo}.} 
\begin{equation}
\dot{Q}_{i}\left(t\right)={\sum_{j}\mu_{ij}\partial_{j}\theta}\Big\vert_{Q\left(t\right)},\label{eq:SchematicVelocitiesGuidingEq}
\end{equation}
 where dots denote time derivatives. Anticipating the Born rule, which
states that the probability density $\rho\left(q,t\right)$ for the
system to be found in its configuration $q$ at the time $t$ should
be given by the modulus-square of the wave function (Born 1926)\nocite{Born:1926zqds},
\begin{equation}
\rho\left(q,t\right)=\Psi\left(q,t\right)\overline{\Psi\left(q,t\right)}=\verts{\Psi\left(q,t\right)}^{2}=R^{2}\left(q,t\right),\label{eq:SchematicBornRule}
\end{equation}
 where bars denote complex-conjugation, it follows from a simple calculation
that the following continuity equation holds: 
\begin{equation}
\partial_{t}\rho=-\sum_{i}\partial_{i}J_{i}.\label{eq:SchematicProbabilityContinuityEq}
\end{equation}
 Here $J_{i}\left(q,t\right)$ is a probability current density given
by 
\begin{equation}
J_{i}=\rho\sum_{j}\mu_{ij}\partial_{j}\theta.\label{eq:SchematicProbabilityCurrentDensity}
\end{equation}
 One can then re-express the guiding equation \eqref{eq:SchematicVelocitiesGuidingEq}
as the statement that the velocity $\dot{Q}_{i}\left(t\right)$ is
equal to the ratio of the corresponding probability current density
$J_{i}$ and the probability density $\rho$, evaluated at the system's
actual instantaneous configuration $Q\left(t\right)=\left(Q_{1}\left(t\right),\dots,Q_{n}\left(t\right)\right)$
at the time $t$: 
\begin{equation}
\dot{Q}_{i}\left(t\right)=\left.\frac{J_{i}}{\rho}\right\vert _{Q\left(t\right)}=\frac{J_{i}\left(Q\left(t\right),t\right)}{\rho\left(Q\left(t\right),t\right)}.\label{eq:GuidingEquationAsRatioCurrentDensityToProbabilityDensity}
\end{equation}
 There are never divide-by-zero problems here because the system's
configuration $Q\left(t\right)$ has, by definition, zero probability
of being located in a region of its configuration space where the
probability density $\rho$ is zero. If the system gets close to a
region of its configuration space with very small probability density,
then its velocities will become highly unstable, driving the system
away from that region. 

Collectively, these equations ensure a crucial feature of the pilot-wave
theory, known today as \textquoteleft equivariance,\textquoteright{}
and which de Broglie originally called ``the principle of interference''
(``\emph{le principe des interférences}''). Equivariance is the
feature that if $\rho\left(q,t_{0}\right)=\verts{\Psi\left(q,t_{0}\right)}^{2}$
is the correct probability density at an initial time $t_{0}$, a
condition known today as the \textquoteleft quantum equilibrium hypothesis,\textquoteright{}
then $\rho\left(q,t\right)=\verts{\Psi\left(q,t\right)}^{2}$ will
continue to be the correct probability density at all later times
$t$. 

\subsection{Bohmian Mechanics\label{subsec:Bohmian-Mechanics}}

In 1951, David Bohm published a new textbook on quantum mechanics,
titled simply \emph{Quantum Theory} (Bohm 1951)\nocite{Bohm:1951qt}.
Chapter 22, ``Quantum Theory of the Measurement Process,'' presented
a new approach to the measurement problem. In particular, in Section
22.8, ``Destruction of Interference in the Process of Measurement,''
Bohm developed an early version of the theory of decoherence. However,
in conversations shortly thereafter, Albert Einstein persuaded Bohm
that decoherence alone was insufficient for solving the measurement
problem.

In 1952, Bohm independently introduced essentially the same pilot-wave
theory as in de Broglie's 1930 book, in a pair of papers both titled
``A Suggested Interpretation of the Quantum Theory in terms of \textquoteleft Hidden
Variables\textquoteright{} '' (Bohm 1952a,b)\nocite{Bohm:1952siqtthvi},
without initially knowing about de Broglie's earlier work (see, for
example, Section 2.3.2, p. 30, in Freire Junior 2015)\nocite{FreireJr:2015tqdrtfoqm11}.
In the acknowledgments section of the first of these two papers, Bohm
wrote, simply, ``The author wishes to thank Dr. Einstein for several
interesting and stimulating discussions.'' (Bohm did not include
an acknowledgments section in his second ``Suggested Interpretation''
paper.) 

At Wolfgang Pauli's prompting,\footnote{From a letter to Pauli dated November 20, 1951: ``I have changed
the introduction of my paper so as to give due credit to de Broglie,
and have stated that he gave up the theory too soon (as suggested
in your letter).'' (Pauli 1996)\nocite{Pauli:1996scwbehao}} Bohm clarified de Broglie's priority in the first ``Suggested Interpretation''
paper: 
\begin{quotation}
After this article was completed, the author's attention was called
to similar proposals for an alternative interpretation of the quantum
theory made by de Broglie in 1926, but later given up by him partly
as a result of certain criticisms made by Pauli and partly because
of additional objections raised by de Broglie himself. {[}Bohm 1952a{]}\nocite{Bohm:1952siqtthvi}
\end{quotation}
Bohm conceded de Broglie's priority explicitly in a letter to Pauli
in the summer of 1951: 
\begin{quotation}
With regard to de Broglie, I am ready to admit that he thought of
the idea first. However, the essential point is not the credit for
who got this idea first. The fact is that de Broglie came to the erroneous
conclusion that the idea does not work. My main contention is that
de Broglie did not carry his ideas to a logical conclusion, and that
if he had, he would have seen that they led to precisely the same
results as are required experimentally and as are obtained from the
usual interpretations. {[}Pauli 1996{]} \nocite{Pauli:1996scwbehao}
\end{quotation}
In a subsequent letter to Pauli in October of 1951, Bohm revisited
the question of priority once again, with a bit of additional metaphorical
flair:
\begin{quotation}
You refer to this interpretation as de Broglie's. It is true that
he suggested it first, but he gave it up because he came to the erroneous
conclusion that it does not work. The essentially new point that I
have added is to show in detail (especially by working out the theory
of measurements in paper II) that this interpretation leads to all
of the results of the usual interpretation. {[}...{]} If one man finds
a diamond and then throws it away because he falsely concludes that
it is a valueless stone, and if this stone is later found by another
man who recognizes its true value, would you not say that the stone
belongs to the second man? I think the same applies to this interpretation
of the quantum theory. {[}Ibid.{]}
\end{quotation}
Bohm's work on decoherence ended up being crucial for clearing up
various subtleties in the pilot-wave theory's approach to resolving
the measurement problem, as Bohm explained, in particular, in Section
2 and in Appendix B of the second of his 1952 ``Suggested Interpretation''
papers: 
\begin{quotation}
As we shall show in Appendix B of Paper II, however, all of the objections
of de Broglie and Pauli could have been met if only de Broglie had
carried his ideas to their logical conclusion. The essential new step
in doing this is to apply our interpretation in the theory of the
measurement process itself as well as in the description of the observer
system. {[}Bohm 1952a{]}\nocite{Bohm:1952siqtthvi}
\end{quotation}
Today this pilot-wave theory is called ``the de Broglie\textendash Bohm
theory,'' or ``Bohmian mechanics.''

Prompted by Einstein and Pauli (Drezet 2023)\nocite{Drezet:2023fftsipwabldbadbsqfaqo},
Bohm contacted de Broglie in 1951 to discuss his theory. News of
Bohm's work had already reached de Broglie, brought to him by de Broglie's
assistant Jean-Pierre Vigier, who would later become one of Bohm's
collaborators.\footnote{As recounted from a 1995 interview of Vigier by David Peat: ``Before
arriving in Brazil, {[}George{]} Yevick had been to Paris to lecture
on Bohm's hidden variable theory. Louis de Broglie was not present,
but later, when his assistant, Jean-Paul Vigier, explained to him
the new development of Bohm's ideas, the older physicist (at least
according to Vigier) became so excited that \textquoteleft he jumped
on a chair in excitement.\textquoteright{} This report convinced de
Broglie to revive his own discarded theory, and he sent Vigier to
Brazil to find out firsthand what was going on.'' (Peat 1997, p.
149)\nocite{Peat:1997iptlatodb}} De Broglie quickly published a short article that same year, titled
``\emph{Remarques sur la Théorie de l'Onde Pilote}'' (``Remarks
on the Pilot-Wave Theory, de Broglie 1951)\nocite{DeBroglie:1951rsltdlop},
explaining why he had abandoned the pilot-wave theory laid out in
his 1930 \emph{Introduction à l'Étude} book. De Broglie wrote:
\begin{quotation}
On the other hand, the pilot-wave theory only veritably attained its
aim, which is a return to an interpretation of wave mechanics conforming
with classical conceptions, if the wave $\Psi$, from which the quantum
potential is derived, is a ``physical reality'' capable of acting
on the movement of the particle or the system, for if it is merely
a representation of the probabilities, as is generally admitted today,
the movement defined by the pilot-wave theory would depend on possibilities
that are not realized, which is paradoxical and takes us completely
away from classical conceptions. However, it seems impossible to consider
the wave $\Psi$ as a physical reality. It is, in effect, represented
by an essentially complex function and, in the case of general systems,
it propagates in configuration space, which is clearly abstract and
fictitious {[}``\emph{l'espace de configuration qui est visiblement abstrait et fictif}''{]}.
{[}De Broglie 1951, in literal English translation from the original
French by the author of the present work{]}\nocite{DeBroglie:1951rsltdlop}
\end{quotation}
De Broglie did not mince words, concluding: 
\begin{quotation}
These considerations and others like them seem to be absolutely opposed
to attributing to the wave $\Psi$ the character of physical reality,
and this still seems to me to be the strongest objection against the
pilot-wave theory. {[}...{]} In summary, the interpretation of wave
mechanics by the pilot-wave theory, to which Mr. Bohm's theory brings
attention, still seems to me to encounter insurmountable difficulties,
principally for the reason of the impossibility of attributing to
the wave $\Psi$ a physical reality or of admitting that the motion
of a particle is determined by possible motions that are not realized.
{[}Ibid.{]}
\end{quotation}

Despite his negative view toward his pilot-wave theory, de Broglie
used the final parts of this article to explain why the pilot-wave
theory evaded von Neumann's famous no-go theorem (von Neumann 1932)\nocite{vonNeumann:1932mgdq}
against the existence of ``hidden parameters'' or ``hidden variables,''
which Bohm took to refer to his particles. Bohm also addressed von
Neumann's no-go theorem himself in a letter to Pauli in October 1951:\footnote{Famously, Grete Hermann and John Bell also found loopholes in von
Neumann's no-go theorem (Hermann 1935; Bell 1966; Crull, Bacciagaluppi
2016)\nocite{Hermann:1935dzinsbs7odngdq,Bell:1966otpohviqm,CrullBacciagaluppi:2016ghbpap}.}
\begin{quotation}
Let us now discuss von Neumann's proof that quantum theory is inconsistent
with hidden causal parameters. His proof involves the demonstration
that no ``dispersionless'' states can exist in the quantum theory,
so that no \emph{single} distribution of hidden parameters could possibly
determine the results of all experiments (including for example, the
measurements of momentum and position). However, von Neumann implicitly
assumes that the hidden variables are only in the observed system
and not in the measuring apparatus. On the other hand, in my interpretation,
the hidden variables are in \emph{both} the measuring apparatus and
the observed system. Moreover, since different apparatus is needed
to measure momentum and position, the actual results in each respective
type of measurement are determined by different distributions of hidden
parameters. Thus, von Neumann\textquoteright s proof is irrelevant
to my interpretation.

It appears that von Neumann has agreed that my interpretation is logically
consistent and leads to all of the results of the usual interpretation.
(This I am told by some people.) Also, he came to a talk given by
me and did not raise any objections. {[}Pauli 1996, emphasis in the
original{]}\nocite{Pauli:1996scwbehao}
\end{quotation}

Bohm took a strong stand on the ontology of the pilot wave, despite
its living in a many-dimensional configuration space. In the first
of his 1952 ``Suggested Interpretation'' papers, received by the
journal \emph{Physical Review} on July 5, 1951, Bohm consistently
referred to the wave function as a ``$\psi$-field,'' in analogy
with the electromagnetic field (Bohm 1952a)\nocite{Bohm:1952siqtthvi}.
He also wrote:
\begin{quotation}
Since the force on a particle now depends on a function of the absolute
value, $R\left(\vec x\right)$, of the wave function, $\psi\left(\vec x\right)$,
evaluated at the actual location of the particle, we have effectively
been led to regard the wave function of an individual electron as
a mathematical representation of an objectively real field. {[}Ibid.{]}
\end{quotation}
However, Bohm had little to say in his two ``Suggested Interpretation''
papers about the ontological meaning of the wave function in the case
of more than one particle, except to say:
\begin{quotation}
In the two-body problem, the system is described therefore by a six-dimensional
Schroedinger wave and by a six-dimensional trajectory, specifying
the actual location of each of the two particles. {[}Ibid.{]}
\end{quotation}
Bohm's second ``Suggested Interpretation'' paper, also received
by the journal \emph{Physical Review} on July 5, 1951, concluded
with these words:
\begin{quotation}
Nevertheless, evidence for the existence of individual atoms was ultimately
discovered by people who took the atomic hypothesis seriously enough
to suppose that individual atoms might actually exist, even though
no one had yet observed them. We may have here, perhaps, a close analogy
to the usual interpretation of the quantum theory, which avoids considering
the possibility that the wave function of an individual system may
represent objective reality, because we cannot observe it with the
aid of existing experiments and theories. {[}Bohm 1952b{]}\nocite{Bohm:1952siqtthvii}
\end{quotation}
In a letter in English to Bohm dated December 3, 1951, Pauli wrote
``I think that the fundamental difference of this $\psi$-function
from classical fields is the fact that the former is in a polydimensional
space and not in ordinary space-time'' (Pauli 1996)\nocite{Pauli:1996scwbehao}.
To this statement, Bohm wrote back the following reply:
\begin{quotation}
Since you admit the logical consistency of my point of view, and since
you cannot give any arguments showing that it is wrong, it seems to
me that your desire to hold on to the usual interpretation can have
only one justification; namely, the positivist principle of not postulating
constructs that do not correspond to things that can not {[}sic{]}
be observed. This is exactly the principle which caused Mach to reject
the reality of atoms, for example, since no one in his day knew how
to observe them. As for your objections about the polydimensional
character of the $\psi$ function, I have shown in my previous letter
that they lead to no inconsistencies and that they provide a perfectly
good set of concepts. There is no reason why some aspects of reality
should not be polydimensional. This merely means a closer unity between
distant systems than we had previously been led to suppose. After
all, we must not expect the world at the atomic level to be a precise
copy of our large scale experience (as proponents of the usual interpretation
are so fond of saying). Rather than accept a perfectly logical and
definite concept of polydimensional reality that leads to the right
results in all known cases, and opens up new mathematical possibilities,
you prefer the much more outlandish idea that there is no way to conceive
of reality at all at the atomic level. Instead you are willing to
restrict your conceptions to results that can be \emph{observed} at
the long scale level, even though more detached conceptions are available,
which show at least, never {[}sic{]} the production of these results
\emph{might} be understood causally and continuously. It is hard for
me to understand such a point of view on the part of a person who
claims that he is not a positivist. {[}Ibid., emphasis in the original{]}
\end{quotation}

Bohm described his formulation of the pilot-wave theory to Einstein,
but it does not appear that Einstein found the theory fully convincing.
In a 1952 letter to Born, dated May 12, 1952, Einstein wrote: 
\begin{quotation}
Have you noticed that Bohm believes (as de Broglie did, by the way,
25 years ago) that he is able to interpret the quantum theory in deterministic
terms? That way seems too cheap to me. But you, of course, can judge
this better than I. {[}Einstein, Born 1971{]}\nocite{EinsteinBornBorn:1971tbelcbaeamahbf1t1wcbmb}
\end{quotation}
Although Pauli was cordial in his letters to Bohm, Pauli was much
more caustic in German-language letters to his colleagues. For example,
in a letter in German to Markus Fierz, dated January 6, 1952, Pauli
wrote:
\begin{quotation}
He {[}Bohm{]} writes me letters like a sectarian priest {[}``\emph{Sektenpfaff}''{]},
to \emph{convert} me \textendash{} and specifically to the old pilot-wave
theory of de Broglie (1926/27), which he has revived. I did suggest
to Bohm that we temporarily suspend our correspondence until he has
new results to report; but that has not helped, as letters from him
come almost daily, often with penalty postage (he apparently has an
unconscious wish to punish me).

{[}...{]}

I now fear that Bohm will also find a considerable number of followers
among young people{[}...{]}

There is naturally no helping a fool {[}``\emph{Narren}''{]} like
Bohm; but do you not think \emph{that such an argument}, to younger
students who want to find their orientation, would make an impression?
Can you think of anything the fools could say against it? (There is
also the danger that \textendash{} if I simply remain silent \textendash{}
Bohm will spread that I have, ``except for philosophical prejudices,''
nothing to object against his ``theory.'') {[}Pauli 1996, emphasis
in the original, in literal English translation from the original
German by the author of the present work{]}\nocite{Pauli:1996scwbehao}
\end{quotation}

\subsection{Everettian Quantum Theory\label{subsec:Everettian-Quantum-Theory}}

These historical debates over the ontological status of the wave function
have had a substantial legacy. Indeed, just a few years after Bohm's
1952 ``Suggested Interpretation'' papers, Hugh Everett introduced
his own new interpretation of quantum theory (Everett 1956, 1957;
Osnaghi, Freitas, Freire Junior 2009)\nocite{Everett:1956ttotuwf,Everett:1957rsfqm,OsnaghiFreitasFreire:2009tooteh},
an interpretation that took Bohm's ontological stance toward the wave
function so seriously that Everett decided that the ``universal wave
function'' was the \emph{only} fundamental ontology needed for the
universe. Everett himself made that monistic claim quite explicitly
in describing the benefits of his interpretation in his unpublished
long-form dissertation in 1956: 
\begin{quotation}
One is thus free to build a conceptual model of the universe, which
postulates only the existence of a universal wave function which obeys
a linear wave equation. {[}Everett 1956, Section VI{]}\nocite{Everett:1956ttotuwf}
\end{quotation}
Everett reiterated this view in the shorter published version of his
interpretation in 1957: 
\begin{quotation}
The wave function is taken as the basic physical entity with \emph{no
a priori interpretation}. Interpretation only comes \emph{after} an
investigation of the logical structure of the theory. {[}Everett 1957,
emphasis in the original{]}\nocite{Everett:1957rsfqm}
\end{quotation}
Everett's theory, of course, is now widely known as the ``many worlds''
interpretation, a name later given to the theory by Bryce DeWitt (Everett,
DeWitt, Graham 1973)\nocite{EverettDeWittGraham:1973mwiqm}.

It is worth highlighting how Everett described Bohm's theory near
the end of that long-form dissertation. Notice in the following passage
how Everett expressed no qualms whatsoever with Bohm's wave-function
realism, how Everett expressly embraced Bohm's reification of the
wave function as a part of the ontology of the world, and also how
Everett made clear that his own theory took the wave function to be
the sole ontology of nature: 
\begin{quotation}
Bohm considers $\psi$ to be a real force field acting on a particle
which always has a well-defined position and momentum (which are the
hidden variables of this theory). The $\psi$-field satisfying Schrödinger's
equation is pictured as somewhat analogous to the electromagnetic
field satisfying Maxwell's equations, although for systems of $n$
particles the $\psi$-field is in a $3n$-dimensional space. With
this theory Bohm succeeds in showing that in all actual cases of measurement
the best predictions that can be made are those of the usual theory,
so that no experiments could ever rule out his interpretation in favor
of the ordinary theory. Our main criticism of this view is on the
grounds of simplicity \textendash{} if one desires to hold the view
that $\psi$ is a real field then the associated particle is superfluous
since, as we have endeavored to illustrate, the pure wave theory is
itself satisfactory. {[}Everett 1956, Section VI.c{]}\nocite{Everett:1956ttotuwf}
\end{quotation}

The timing of Everett's work could hardly have been a coincidence\textemdash Everett
started graduate school in physics at Princeton in 1953. Bohm had
been an assistant professor at Princeton until his departure in 1951.
When Everett arrived, Bohm's textbook and his ``Suggested Interpretation''
papers were still brand-new. Bohm's 1951 textbook and his two 1952
``Suggested Interpretation'' papers were the first three cited works
in Everett's 137-page long-form dissertation, which included fewer
than two dozen citations altogether. In fact, Bohm's textbook was
the \emph{only} pedagogical textbook on quantum mechanics cited by
Everett in his dissertation, and Everett's detailed footnotes clearly
indicated that he had carefully read Bohm's later chapters on the
measurement process and Bohm's original treatment of what would now
be called decoherence, all of which played prominent roles in Everett's
dissertation. 

Schrödinger's writings also clearly had an influence on Everett. Here
is another way in which Everett described his interpretation in his
long-form dissertation: 
\begin{quotation}
e. \emph{The wave interpretation}. This is the position proposed in
the present thesis, in which the wave function itself is held to be
the fundamental entity, obeying at all times a deterministic wave
equation.

This view also corresponds most closely with that held by Schrödinger.
{[}Everett 1956, Section VI.e{]}\nocite{Everett:1956ttotuwf}
\end{quotation}
The last sentence in this passage included a footnote to the first
of a pair of 1952 papers by Schrödinger titled ``Are There Quantum
Jumps?'' (Schrödinger 1952a,b)\nocite{Schrodinger:1952atqjpi,Schrodinger:1952atqjpii}.
In these two papers, Schrödinger returned to arguing that the waves
of quantum theory were physically real, expressed the view that perhaps
only waves were needed for a physical description of nature, and explicitly
questioned the need for separate probability postulates:
\begin{quotation}
After this has been recognised, is the probability scheme any longer
needed? Has the idea of the mysterious sudden leaps of single electrons
not become gratituitous? Is it expedient? The waves are there anyhow,
and we are not at a loss to prove it. {[}Schrödinger 1952a{]}\nocite{Schrodinger:1952atqjpi}
\end{quotation}

How did Schrödinger assuage his earlier concerns about the physical
reality of waves propagating in a many-dimensional configuration space?
In the second of his two 1952 papers, Schrödinger cited an earlier
1950 paper, titled ``What is an Elementary Particle?'' (Schrödinger
1950)\nocite{Schrodinger:1950wiaep} in which he argued that second
quantization had provided an escape hatch: 
\begin{quotation}
A method of dealing with the problem of many particles was indicated
in 1926 by the present writer. The method uses waves in many-dimensional
space, in a manifold of $3N$ dimensions, $N$ being the number of
particles. Deeper insight led to its improvement. The step leading
to this improvement is of momentous significance. The many-dimensional
treatment has been superseded by so-called second quantization, which
is mathematically equivalent to uniting into one three-dimensional
formulation the cases $N=0,1,2,3,...$ (to infinity) of the many-dimensional
treatment. This highly ingenious device includes the so-called new
statistics, with which we shall have to deal below in much simpler
terms. It is the only precise formulation of the views now held, and
the one that is always used. {[}Ibid.{]}
\end{quotation}
It would be an understatement to say that this claim did not settle
the controversy over the physical reality of the wave function or
quantum state. After all, for second-quantized systems, such as those
used in quantum field theory, the quantum state does not go away or
become less abstract. 

Recall that for a nonrelativistic system consisting of a fixed number
of finitely many particles, the system's quantum state can be represented
as a wave function $\Psi\left(q_{1},\dots,q_{n},t\right)$ that propagates
from one time $t$ to another in an $n$-dimensional configuration
space. Analogously, for a bosonic quantum field, the quantum state
can be represented as a wave functional $\Psi\left[\left\{ \phi\right\} ,\Sigma\right]$
that propagates from one spacelike hypersurface $\Sigma$ to another
in the configuration space of the field $\phi$, where the dimension
of that configuration space is continuously infinite. (See, for instance,
Sebens 2022\nocite{Sebens:2022tfof}.) Although it is true that the
local field operators of a quantum field are assigned to points in
four-dimensional spacetime, it is important to remember that local
field operators do not define the quantum state of the quantum field
as a whole, any more than the algebra of observables of a generic
quantum system define the system's quantum state as a whole. Furthermore,
one should be careful not to conflate either the quantum state or
the local field operators of a quantum field with the local magnitudes
of a classical field. 

\section{Conclusion\label{sec:Conclusion}}

Debates over the ontological nature of the wave function continue
to this day, inspiring approaches to quantum foundations like the
Harrigan-Spekkens ``ontological models'' framework (Harrigan, Spekkens
2010)\nocite{HarriganSpekkens:2010eievqs}. One of the most prominent
recent theorems in quantum foundations, due to Matthew Pusey, Jonathan
Barrett, and Terry Rudolph, and known widely as the \textquoteleft PBR
theorem,\textquoteright{} was also aimed squarely at the question
of the ontological status of the wave function. The original 2012
PBR paper opened its introduction with words that should sound very
familiar by this point:
\begin{quotation}
At the heart of much debate concerning quantum theory lies the quantum
state. Does the wavefunction correspond directly to some kind of physical
wave? If so, it is an odd kind of wave, as it is defined on an abstract
configuration space, rather than the three-dimensional space in which
we live. {[}Pusey, Barrett, Rudolph 2012{]}\nocite{PuseyBarrettRudolph:2012rqs}
\end{quotation}
Due to limitations of space, nothing more will be said about these
contemporary developments in the present work.

\section*{Acknowledgments}

The author would especially like to thank David Albert, Branden Fitelson,
Barry Loewer, and Tim Maudlin for helpful discussions.

\bibliographystyle{1_home_jacob_Documents_Work_My_Papers_2026-Agai___n_in_Bohmian_Mechanics_custom-abbrvalphaurl}
\bibliography{0_home_jacob_Documents_Work_My_Papers_Bibliography_Global-Bibliography}

\end{document}